\documentstyle[twoside,fleqn,epsfig]{article}

\def\Journal#1#2#3#4{{#1} {#2} (#3) #4 }

\def\APP{\em Astrop. Phys.}

\def\PLB{{\em Phys. Lett.} B}

\def\PRD{{\em Phys. Rev.} D}

\def\be{\begin{equation}}
\def\ee{\end{equation}}
\def\bea{\begin{eqnarray}}
\def\eea{\end{eqnarray}}
\newcommand{\lsim}{\mathrel{\mathop{\kern 0pt \rlap
  {\raise.2ex\hbox{$<$}}}
  \lower.9ex\hbox{\kern-.190em $\sim$}}}
\newcommand{\gsim}{\mathrel{\mathop{\kern 0pt \rlap
  {\raise.2ex\hbox{$>$}}}
  \lower.9ex\hbox{\kern-.190em $\sim$}}}

\newcommand{\AmS}{{\protect\the\textfont2
  A\kern-.1667em\lower.5ex\hbox{M}\kern-.125emS}}
\normalsize
\begin{document}

\begin{flushright}
{\bf ROM2F/2005/19\\} 
{\bf Int. J. Mod. Phys. A (in press)\\}
\end{flushright}
\vspace{-0.4cm}

\normalsize

\baselineskip=0.65cm
\vspace*{0.5cm}

\begin{center}
\Large \bf
Investigating pseudoscalar and scalar dark matter \\
\rm
\end{center}

\vspace{0.5cm}
\normalsize

\noindent \rm R.\,Bernabei,~P.\,Belli,~F.\,Montecchia,~F.\,Nozzoli

\noindent {\it Dip. di Fisica, Universita' di Roma "Tor Vergata"
and INFN, sez. Roma2, I-00133 Rome, Italy}

\vspace{3mm}

\noindent \rm F.\,Cappella, A.\,Incicchitti,~D.\,Prosperi

\noindent {\it Dip. di Fisica, Universita' di Roma "La Sapienza"
and INFN, sez. Roma, I-00185 Rome, Italy}

\vspace{3mm}
                      
\noindent \rm R.\,Cerulli

\noindent {\it Laboratori Nazionali del Gran Sasso, I.N.F.N., Assergi, Italy}

\vspace{3mm}

\noindent \rm C.J.\,Dai,~H.L.\,He,~H.H.\,Kuang,~J.M.\,Ma,~Z.P.\,Ye\footnote{also:
University of Jing Gangshan, Jiangxi, China}

\noindent {\it IHEP, Chinese Academy, P.O. Box 918/3, Beijing 100039, China}

\vspace{0.5cm}
\normalsize

\begin{abstract}

In this paper another class of Dark Matter candidate particles:
the pseudoscalar and scalar light bosonic candidates, is discussed. Particular care is devoted  
to the study of the processes for their detection 
(which only involves electrons and photons/X-rays)
in a suitable underground experimental set-up.
For this purpose the needed calculations are developed and
various related aspects and phenomenologies are discussed.
In particular, it is shown that -- in addition to the WIMP cases already discussed elsewhere     
-- there is also possibility for a bosonic candidate to account for the
6.3 $\sigma$ C.L. model independent evidence for the presence of a particle DM component in   
the galactic halo observed by DAMA/NaI. Allowed regions in these scenarios are presented also 
paying particular care on the cosmological interest of the bosonic candidate.

\end{abstract}



\section{Introduction}
 
In order to investigate in a model independent way the presence of a DM particle component in the galactic halo
we have exploited the effect of the Earth revolution around the Sun on the DM particles
interactions on suitable underground detectors. In fact, as a consequence of its
annual revolution, the Earth should be 
crossed by a larger flux of DM particles in June (when its rotational velocity is summed 
to the one of the solar system with respect to the Galaxy)
and by a smaller one in December (when the two velocities are subtracted).
This offers an efficient model independent signature, able to test a large interval of 
cross sections and of halo densities; it is named {\it annual modulation 
signature} and was originally suggested in the middle of '80 by \cite{Freese}. 
The annual modulation signature is very distinctive
since a DM particle  induced seasonal effect must simultaneously satisfy
all the following requirements: the rate must contain a component
modulated according to a cosine function (1) with one year period (2)
and a phase that peaks roughly around $\simeq$ 2$^{nd}$ June (3);
this modulation must only be found in a well-defined low energy range, 
where DM particles induced events can be present  (4); it must apply to those events in
which just one detector of many actually "fires", since
the dark matter particle multi-scattering probability is negligible (5); the modulation
amplitude in the region of maximal sensitivity must be $\lsim$7$\%$ for usually adopted halo distributions (6), but it can
be larger in case of some possible scenarios such as e.g. those in refs. \cite{Wei01,Fre04,Sikivie}.
Only systematic effects able to fulfil these 6 requirements and to account for the whole observed modulation amplitude 
could mimic this signature; thus, no other effect investigated so far in the field of rare processes offers 
a so stringent and unambiguous signature.

The DAMA/NaI set-up has exploited such a DM annual modulation signature over seven annual cycles
\cite{Prop,Mod1,Mod2,Ext,Mod3,Sist,Sisd,Inel,Hep,RNC,IJMPD}. It has achieved 
a 6.3 $\sigma$ C.L. model independent 
evidence for the presence of a DM particle component in the galactic halo. 
Some of the many possible corollary quests for the candidate particle have been carried out mainly focusing  
so far the class of DM candidate particles named WIMPs \cite{Mod1,Mod2,Ext,Mod3,Sist,Sisd,Inel,Hep,RNC,IJMPD}.
In literature several candidates for WIMPs 
have been considered -- all foreseen in theories beyond the Standard Model of particle Physics --
such as the neutralino in the supersymmetric theories \cite{Bo03,Bo04,Botdm}, a subdominant 
sneutrino (the spin-0 supersymmetric partner of the neutrino), a subdominant neutrino 
of the fourth family (see the analysis including the results of the indirect search
in ref. \cite{khlopov}), a dominant sneutrino in supersymmetric models with 
violation of lepton number, where two mass states and a small energy splitting
is present, as reported in ref. \cite{Wei01},
the particles from multi-dimensional Kaluza-Klein-like theories, etc.
In addition, other possibilities still exist with a phenomenology
similar as for the WIMP cases: 
the mirror Dark Matter particles \cite{foot},
the self-interacting dark matter particles \cite{Saib}, etc.
Moreover, in principle even whatever 
particle with suitable characteristics, not yet foreseen by theories, can be a good candidate as 
DM in the galactic halo.

In the present paper the class of light bosonic candidates -- either with the pseudoscalar
or with the scalar coupling -- will be investigated 
in some of the possible scenarios.
Moreover, also here we will not yet consider either the case of halos with caustics or 
with streams contributions \cite{Sikivie,Fre04}, which can have significant impact 
for all the DM candidate particles; we plan to exploit these scenarios
in near future by further devoted analyses.

It is worth to note that the direct detection process for light bosonic DM candidates is 
based on the total conversion in NaI(Tl) crystal of the mass of the absorbed bosonic particle 
into electromagnetic radiation. Thus, in these 
processes the target nuclei recoil is negligible and is not involved in the detection 
process; therefore, signals from these light bosonic DM candidates 
are lost in experiments based on rejection procedures of the electromagnetic contribution
to the counting rate.

\section{The light bosons candidates and their main interactions with ordinary matter}

Let us firstly remind the Peccei-Quinn (PQ) axion; it is a pseudoscalar particle hypothesized 
in order to solve the strong CP problem. Several models exist which describe the coupling 
of such a particle to ordinary 
matter; the most popular ones are the DFSZ, where the particle is directly coupled with electron \cite{DFSZ}, and the 
KSVZ, where such a coupling is not present at tree-level \cite{KSVZ}.
However, many models exist for axion-like particles,
that is particles having similar phenomenology with ordinary matter as the axion, but
which allow values for the coupling constants and for the mass 
significantly different from those foreseen in the DFSZ and KSVZ models. For 
example, we mention the axion itself in the Kaluza-Klein theories \cite{kkax},
where it would have similar couplings as in the 
DFSZ and KSVZ models, but  much higher mass states or the "exotic" axion models proposed by \cite{exax}.
Other candidates 
are pseudo-Nambu-Goldstone bosons related to spontaneous global symmetry breaking 
different from the U(1)$_{PQ}$ hypothesized
by Peccei-Quinn, such as the pseudoscalar familon in the case of the family symmetry 
or the Majoron for the lepton number symmetry \cite{RG,Majo}. 

The previous cases mainly refer to light bosonic candidates with pseudoscalar coupling;
obviously, cases with scalar coupling can be considered as well,
such as e.g. the scalar familon \cite{Rie82} and the sgoldstino \cite{sgold}.

Moreover, it is also interesting to remark that some indirect astrophysical observations:
i) the Solar corona problem;
ii) the X-rays flux detected by ROSAT in the direction of the dark side of the Moon;
iii) the X-rays background radiation in the 2-8 keV region measured by CHANDRA (XRB);
iv) the excess of X-rays from clusters of galaxies;
have recently been analysed in a model of 
axion-like particles with mass in the keV range and coupling to photons
$g_{a \gamma \gamma}$ of the order of $10^{-13} GeV^{-1}$ \cite{DiLella}, that is requiring a model 
with expectations for the coupling constants and masses well different with the respect to
those expected in the DFSZ and KSVZ models. 

It has also been argued that the existence of axion-like particles may account for the high
energy cosmic rays \cite{Gorb01}. Also in this case the bosonic candidate particle
can have masses of few keV and $g_{a \gamma \gamma}$ of the order of $10^{-15}$ 
GeV$^{-1}$, that is still couplings well different than those expected by the 
DFSZ and KSVZ models.

Finally, a keV Majoron has been suggested as DM particle \cite{Majo2} and
a $\sim$ keV DM pseudoscalar candidate has also been taken into account in ref. \cite{exax,DMpse}.

In this paper, we consider a light bosonic candidate, either with pseudoscalar or with scalar 
couplings, of $\sim$ keV mass as DM component in the galactic 
halo\footnote{In fact, considering the phenomenology of this candidate (see later),
a keV-scale bosonic candidate naturally arises as an additional solution 
to the observed model-independent DAMA/NaI annual modulation signal.}. Several mechanisms can be advocated for
the production of these particles in the early Universe (see e.g. ref. \cite{Majo2,exax,DMpse});
in particular, it has been demonstrated that these particles can be of cosmological interest, providing
many configurations with $\Omega \sim 0.3$. Moreover, such a DM keV candidate can also be able to explain 
the galactic scale in the structure formation problem \cite{Majo2}.

\begin{figure} [!ht]
\centering
\includegraphics[width=350pt] {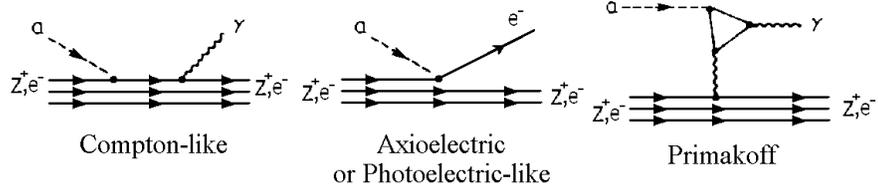}
\caption{Possible diagrams for the direct detection processes of a light boson particle.
{\em On the left}: ``Compton - like'' effect;
{\em on the center}: ``Axioelectric'' or photoelectric-like effect;
{\em on the right}: Primakoff effect. See text.}
\label{fg:diagr}
\end{figure}

The diagrams of Fig. \ref{fg:diagr} show the main processes involved in the detection of 
a DM light bosonic particle (here generically named $a$)
both in the pseudoscalar and in the scalar interaction types.

As first, let us recall that in the case of interest here 
the DM light bosonic candidates are non-relativistic 
since they should be trapped in the galactic 
halo\footnote{Moreover, in case these particles are relics 
from the early Universe, they should have been cooled down by the cosmological 
redshift; thus, they can be non-relativistic now, similarly as the case of 
massive neutrinos. Other scenarios for their production and for their
non-relativistic properties can be in principle considered as well.
On the other hand, in order to induce the observed 
annual modulation signal, they cannot obviously be relativistic (see also later).}.

On the left of Fig. \ref{fg:diagr} the ``Compton - like'' effect is shown; 
in fact, considering that the light boson particle couples to charged fermions of the ordinary matter 
(either electrons or $u$ and $d$ quarks), 
it is absorbed as in the ordinary Compton effect and 
it is emitted a photon, whose energy is $\simeq m_a$ 
if -- as in the case of interest here -- 
$m_a$ is much lower that the charged fermions masses.

In the center of Fig. \ref{fg:diagr} the ``axioelectric effect'' or photoelectric-like effect 
is shown; there the $a$ particle -- in the hypothesis of a non-zero coupling to electrons --
is absorbed by a bound electron of the atom exciting or ionizing it. 
Thus, the sum of the kinetic energy of the photoelectron and 
of the energy produced by X-rays and Auger electrons in the re-arrangement of the atomic shell 
is $\simeq m_a$; this is the detected quantity.
This effect is phenomenological
similar to the ordinary photoelectric effect. 

On the right of Fig. \ref{fg:diagr} the Primakoff effect is shown; in this case $a$ is coupled with two photons
through a charged fermions loop (leptons or quarks). One of the two photons is virtual and is related 
to the electric field of the atoms in the NaI(Tl) crystal, the other photon has energy $\simeq m_a$
and can be detected.

For the sake of completeness, we note that the elastic scatterings of these light bosonic particles 
on atoms, nuclei and electrons are not considered here since their effect is negligible in underground
direct detection experiments.

In all the processes described above the total (including the secondary processes: X-rays and Auger electrons)
energy release, $E_{rel}$, in the detector (providing that
its detection efficiency is $\simeq 1$ for low-energy electrons and low-energy photons)
matches the total energy of the $a$ particle, $E_a \simeq m_a$ since the $a$ velocity is
of the order of $10^{-3}c$. 
Therefore, the differential cross section can generally be written as:

\begin{equation}
\frac{d\sigma_{tot}}{dE_{rel}} = \sigma_{tot} (v) \delta(E_{rel}-E_a) \simeq
\sigma_{tot} (v) \delta(E_{rel}-m_a)
\label{eq:rategen}
\end{equation}
where $\sigma_{tot}$ generally depends on the DM particle velocity in the
laboratory frame, $|\vec{v}| = v$.
  
Taking into account the energy resolution of the detectors, 
$\Delta$, 
the total differential counting rate per unit target 
as function of the detected energy,
$E_{ee}$, is:

\begin{equation}
\resizebox{0.85\textwidth}{!}{
$\frac{dR_{tot}}{dE_{ee}} =
\int dE_{rel}
\frac{e^{ 
-\frac{(E_{ee}-E_{rel})^2} {2 \Delta^2} }}
{\sqrt{2\pi}\Delta}
\int d^3v \frac{d\phi(\vec{v})} {d^3v}
\frac{d\sigma_{tot}} {dE_{rel}} =
R_{tot} \times
\frac{1}
{\sqrt{2\pi}\Delta} e^{ 
-\frac{(E_{ee}-m_a)^2} {2 \Delta^2}
}
$}
\label{eq:ratecomg}
\end{equation}
where $R_{tot}$ is the area of the gaussian peak centered at the $a$ particle mass.

The incoming flux of $a$ particles is given by $d\phi(\vec{v})=\frac{\rho_a v} {m_a} 
f(\vec{v}+\vec{v}_{\oplus}) d^3v$, with $\rho_a = \xi \cdot \rho_{halo}$ 
local density (in GeV $\cdot$ cm$^{-3}$) of the $a$ particles in the galactic halo ($\rho_{halo}$ is the local halo density and 
$\xi \le 1$ is the $a$ particle fraction amount of the local density).
In the following for simplicity we will assume $\xi = 1$, since the scaling of the presented 
results can be derived straightforward for the other cases. 
Moreover, $f(\vec{v}+\vec{v}_{\oplus})$ is the DM particle velocity distribution, which also 
depends on the Earth velocity in the galactic frame, $\vec{v}_{\oplus}$, 
and, thus, on the time along the year. 
Therefore, the total counting rate per unit target is:
\begin{equation}
R_{tot} = \int d^3v \frac{\rho_a v} {m_a} f(\vec{v}+\vec{v}_{\oplus}) \sigma_{tot}(v)
\label{eq:ratecomtot}
\end{equation}
and, generally, it can depend on the time along the year. 
As an example, we cite here a case, that we will see again in the following:
let us assume that $\sigma_{tot}(v) \propto v$; thus:
\begin{equation}
R_{tot} \propto \int d^3v \; v^2 f(\vec{v}+\vec{v}_{\oplus}) = \langle v^2 \rangle 
\label{eq:ratecomtot2}
\end{equation}
Defining the velocity of $a$ particle in the galactic frame as 
$\vec{v}_g = \vec{v}+\vec{v}_{\oplus}$, we obtain for non rotating halo:
$\langle v^2 \rangle = \langle v_g^2 \rangle + v_{\oplus}^2 $.
Since the Earth velocity in the galactic frame, $\vec{v}_{\oplus}$, is given by the sum of the
Sun velocity, $\vec{v}_{\odot}$, and of the Earth's orbital time-dependent velocity
around the Sun, $ \vec{v}_{SE}(t)$, we obtain neglecting the $v^2_{SE}$ term:

\begin{equation}
\langle
v^2
\rangle \simeq
\langle
v_g^2
\rangle+
v_{\odot}^2
+ 2 \vec{v}_{\odot} \cdot \vec{v}_{SE}(t) \simeq
\langle
v_g^2
\rangle+
v_{\odot}^2
+ v_{\odot} v_{SE} cos (\omega(t-t_0))
\label{eq:vel32}
\end{equation}

In the last equation we have considered the angle of $\sim 60^o$
of the terrestrial orbit with the respect to the galactic plane,
$\omega=2 \pi/T$ with $T=1$ year and the phase $t_0$ corresponding 
to $\simeq$ 2nd June (when the Earth's speed in the galactic frame is
at the maximum).
The Sun velocity can be written as $ |\vec{v}_{\odot}| \simeq v_0 + 12 $ km/s,
where $v_0$ is the local velocity, whose value is in the range 170-270 km/s
\cite{Ext,loc}; the Earth's orbital velocity is $ v_{SE} \simeq 30 $ km/s.
Moreover, $\langle v_g^2 \rangle$ depends on the halo model and on
the $v_0$ value (just for reference, in the particular unphysical case of
isothermal halo model:  $\langle v_g^2 \rangle = \frac{3}{2} v_0^2$).
Finally, in this illustrative case $R_{tot} = S_0 + S_m \cdot cos (\omega(t-t_0))$,
giving a time dependent contribution with amplitude depending on
the adopted halo model.  

In the following, the contributions of each process and of each coupling
to the $\sigma_{tot}(v)$ will be shown, as well as the contributions to $R_{tot}$ 
through the eq. (\ref{eq:ratecomtot}). Note that in all the following 
formulae we assume $\hbar=c=1$.
For clariteness, hereafter the DM bosonic candidate particle 
will be named either $a$ if it has a pseudoscalar interaction or $h$ if it has a scalar
one.

\section{The pseudoscalar case} \label{sec3} 

\begin{figure} [!ht]
\centering
\includegraphics[width=300pt] {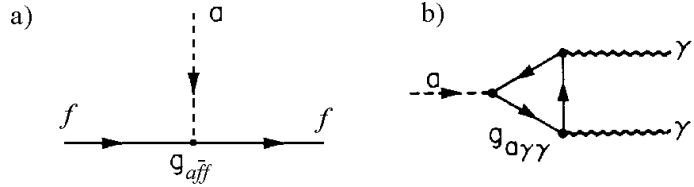}
\caption{Coupling of the pseudoscalar $a$ particle to the fermions $f$ (a case) 
and to the photons through the loop of charged fermions (b case).}
\label{fg:diagramf}
\end{figure}

In Fig. \ref{fg:diagramf} the diagrams, which describe the interactions 
of the pseudoscalar $a$ particle with the fermions $f$ 
(in general leptons and quarks) and the 
"anomalous" coupling to two photons, are shown.

The effective Lagrangian of interaction of the pseudoscalar $a$ particle is:
\begin{equation}
L_{int} = i \;g_{a \bar{f}f} a \; \bar{f} \gamma_5 f + 
\frac{g_{a \gamma \gamma}}{4}a \; F_{\mu \nu}\tilde{F}^{\mu \nu}
\label{eq:laff}
\end{equation}
where the second term describes the effective coupling to the photons through
a charged fermion loop.
As a consequence the pseudoscalar $a$ particle can decay 
into fermions and photons.
The decay in couples $\bar{f}f$ could obviously be possible if 
$a$ would have a mass larger than two times the fermion mass, $m_f$. 

In order to be of cosmological interest as DM candidates in the Universe
the lifetime ($\tau_a$) of $a$ should roughly be of the order or larger than the age 
of the Universe\footnote{The depletion of these particles over 
the course of the history of the Universe is largely dominated by the decay processes.
Therefore, the effect of depletion through conversion processes on ordinary matter 
are consequently not addressed here.},
although this condition might be released in case other exotic scenarios would be considered. 
Its lifetime can be derived considering just the second term of eq. (\ref{eq:laff}) under the assumptions that
$m_a < 2 m_e \simeq 1$ MeV and that
the possible decay into massive neutrinos (whose coupling constants $g_{a \bar{\nu}\nu}$ can
be expected to be very low)
would be neglible with the respect to the two-photon decay contribution:
\begin{equation}
\tau_a = \frac{64 \pi}{g_{a \gamma \gamma}^2 m_a^3}
\label{eq:ta}
\end{equation}

The effective coupling constant to photons
$g_{a \gamma \gamma} $ 
can be calculated by taking into account the charged fermions loop with $Q_f$ charge.
For a pseudoscalar particle having $ m_a << m_f$ one gets \cite{coupl00}:
\begin{equation}
g_{a \gamma \gamma}= \sum_f \frac {g_{a \bar{f}f} Q_f^2 \alpha } {\pi m_f}
\left[1+\frac{m^2_a}{12m^2_f} 
+O \left(\frac{m^4_a}{m^4_f}\right) \right]
\label{eq:triangolos}
\end{equation}

\subsection{Direct detection of the pseudoscalar particle} 

\subsubsection{The Compton-like effect } \label{3.1} 

Let us take into account the diagram of Fig. \ref{fg:diagr}, where
a pseudoscalar particle having a momentum $\vec{k}=m_a \vec{v}$
is absorbed by a charged free and point-like fermion and a photon is emitted.
In this case, in the non-relativistic limit, one gets (see Appendix):
\begin{equation}
\sigma_{C-l,f} \simeq
\frac{\alpha Q_f^2 g^2_{a \bar{f}f} m_a^2}    {2 m_f^4 |\vec{v}|}
\label{eq:sig3}
\end{equation}
considering that the fermion recoil is negligible.

Since the pseudoscalar interaction gives a spin-dependent contribution
(see Appendix), the coherent conversion on closed (spinless) shells vanishes.
Hence, the counting rate $R_{C-l}$ expected in a detector, 
made of various species of target fermions 
(electrons and nuclei in NaI(Tl)) identified with the $f$ index, is:
\begin{equation}
R_{C-l} =
N_T \sum_f 
\frac{\rho_a \alpha Q_f^2 g^2_{a \bar{f}f} m_a} {2 m_f^4} F_f(q)
\label{eq:ratecom}
\end{equation}
where $N_T$ is the number of target atoms (either Na or I atoms in NaI(Tl) detectors)
and the sum runs over all the electrons 
and over the nuclei (if they have unpaired nucleon, as it is the case 
for $^{23}$Na and $^{127}$I nuclei).
$F_f$ is the incoherent scattering function \cite{shi65} of the transferred momentum,  
$q \simeq m_a$. 

As far as regards electrons, the incoherent scattering function 
takes into account their binding in the atoms. 
In particular, a good approximation for non relativistic $a$ particle is to consider
that $F_f(m_a)=1$ when the electron recoil energy ($\simeq \frac {m_a^2} {2 (m_e+m_a)}$) is larger then the binding
energy of the bound electron, zero elsewhere.
For this purpose, we report in Tab. \ref{tb:shell} the
binding energies for the electrons in Na and I atoms.

\begin{table}[!ht]
\caption{Binding energies of elettrons in $^{23}Na$ and $^{127}I$ \cite{Leder}
}
\begin{center}
\begin{tabular}{|c|c|c|}
\hline \hline
Shell  & $E_{Na}$ (eV) & $E_{I}$ (eV) \\
\hline
 1s    &       1070.8 &           33169 \\     
 2s    &       63.5 &           5188 \\     
$ 2p_{1/2} $   &       30.4 &     4852 \\     
$ 2p_{3/2}  $  &       30.5 &     4557 \\     
 3s    &     few eV 
 &     1072 \\     
$ 3p_{1/2}$    &     -  &     931 \\     
$ 3p_{3/2}  $  &     -  &     875 \\     
$ 3d_{3/2}   $ &     -  &     630.8 \\     
$ 3d_{5/2}   $ &     -  &     619.3 \\     
 4s    &     -  &     186 \\     
$ 4p_{1/2} $   &     -  &     123 \\     
$ 4p_{3/2} $   &     -  &     123 \\     
$ 4d_{3/2} $   &     -  &     50.6 \\     
$ 4d_{5/2} $   &     -  &     48.9 \\     
$ 5s  $   &     -  &     few eV \\     
$ 5p  $   &     -  &     few eV \\     
\hline \hline
\end{tabular}
\end{center}
\label{tb:shell}
\end{table}

For the case of nuclei, eq.(\ref{eq:ratecom}) still holds when using for $Q_f$ the
effective charges of the nuclei, $Z_{eff}$, (which take into account
the screening of the electrons) and for the coupling constants 
the following scaling law (valid for nuclei with unpaired proton, as Na and I are):

\begin{equation}
\frac{g_{a \bar{A}A}} {m_A} \sim
\frac{g_{a \bar{p}p}} {m_p} \simeq 
\frac{1}{3} \left( 4 \frac {g_{a \bar{u}u}} {m^*_u}
- \frac{g_{a \bar{d}d}} {m^*_d} \right)
\label{eq:alele9}
\end{equation}
where $m^*_u$ and $m^*_d$ are the constituent up and down quark masses. 
This scaling law has been determined with a similar procedure than those
used for the evaluation of nucleon magnetic moments in NRQM \cite{NRQM}.
The incoherent scattering function for the nuclei can be considered as one 
for the energy of interest. 
As it can be seen, the contribution of the nuclei to the total cross section 
can be generally neglected because of the larger mass of the nuclei with the respect to 
the electron mass.

Finally, it is worth to note that, due to the dependence of $\sigma_{C-l,f}$ on the inverse of the 
$a$ velocity,
in this particular case $R_{C-l}$ does not depend on the time along the year ($S_m=0$).
Therefore, this process only contributes to the constant part, $S_0$, of the signal.

\subsubsection{The axioelectric effect}
\label{a-e}

The process similar to the usual photoelectric effect has been investigated.
Hence, it has been calculated (see Appendix) 
the differential cross section for the process of absorption of 
the pseudoscalar $a$ particle (having a momentum $\vec{k}$) by 
an atomic electron (described by the wave function $\psi_{n l m}$) 
which is extracted (with final momentum $\vec{p}$) from the atomic shell 
with quantum numbers $nlm$:

\begin{equation}
\frac{d\sigma_{A-e,nlm}}{d\Omega} \simeq
g^2_{a\bar{e}e}
\frac{\pi |\vec{k}|}{2 m_e} 
\frac{|\vec{p}|}{(2 \pi)^3} 
\left| \int e^{-i(\vec{p}-\vec{k})\vec{x}} \psi_{n l m} d^3x \right|^2 
\label{eq:aFGR2}
\end{equation}

Therefore, since in this case $\vec{q}=\vec{p}-\vec{k} \simeq \vec{p}$, we expect
$\sigma_{A-e,nlm} \propto v$ and 
a non-zero modulation term appears, as described before.  
To evaluate the $\psi_{n l m}$ Fourier transform integral,
we have approximated the wave functions to the hydrogenic case: 
$ \int e^{-i\vec{q}\cdot\vec{x}} \psi_{n l m} d^3x = (2 \pi)^{3/2}
F_{n l} (q) \cdot Y_{l m} (\Omega)$, where $Y_{l m}$ are the spherical harmonics.
Defining $Q=\frac{|\vec{q}|}{Zq_0}$ with $q_0=1/a_0$ 
and $a_0$ Bohr radius, one gets
\cite{ppa}:

\begin{equation}
\resizebox{0.85\textwidth}{!}{
$F_{n l}(q)=\left[\frac{2(n-l-1)!}{\pi(n+l)!} \right]^{1/2} 
n^22^{2l+2}l! \frac{n^lQ^l}{(n^2Q^2+1)^{l+2}}
C^{l+1}_{n-l-1}\left(\frac{n^2Q^2-1}{n^2Q^2+1}\right)
(Zq_0)^{-3/2}$ 
}
\label{eq:hfgen}
\end{equation}

where $C^{\alpha}_{j}(x)$, the Gegenbauer polynomials,
are defined by the serie:

\begin{equation}
(1-2xs+s^2)^{-\alpha}=\sum_jC^{\alpha}_{j}(x)s^j
\label{eq:hserie}
\end{equation}

Then, one obtains:

\begin{equation}
C^{\alpha}_{0}(x)=1
\label{eq:c0}
\end{equation}

\begin{equation}
C^{\alpha}_{1}(x)=2\alpha x
\label{eq:c1}
\end{equation}

\begin{equation}
C^{\alpha}_{2}(x)=2\alpha(\alpha+1)x^2-\alpha
\label{eq:c2}
\end{equation}

\begin{equation}
C^{\alpha}_{3}(x)=\frac{4}{3}\alpha(\alpha+1)(\alpha+2)x^3-2\alpha(\alpha+1)x
\label{eq:c3}
\end{equation}

\begin{equation}
C^{\alpha}_{4}(x)=\frac{2}{3}\alpha(\alpha+1)(\alpha+2)(\alpha+3)x^4
                  -2\alpha(\alpha+1)(\alpha+2)x^2
                  +\frac{1}{2}\alpha(\alpha+1)
\label{eq:c4}
\end{equation}

From these relations it is possible to obtain all the 
$F_{n l}$ for the Iodine and Sodium atoms.

Let us now consider the counting rate $R_{A-e}$ expected in a detector
made by various kinds of target atoms -- as it is the case of NaI(Tl) -- 
identified by the $b$ index:
\begin{equation}
R_{A-e} =
N_T
\frac{\rho_a \pi g^2_{a\bar{e}e}} {2m_e}
\langle v^2 \rangle
\sum_{b=Na,I} \sum_{nl} N_{nl}
F^2_{nl}(q) \; p \;
\Theta(m_a-E_{nl})
\label{eq:sratecomae2}
\end{equation}
with $q \simeq p \simeq \sqrt{2m_e(m_a-E_{nl})}$. 
There the second sum is running over $nl$ levels -- each of them with $N_{nl}$ electrons -- allowed
by the Heaviside step function $\Theta(m_a-E_{nl})$.
The binding energy $E_{nl}$ of Na or I atoms have already been 
reported in Tab. \ref{tb:shell}.

\subsubsection{The Primakoff effect}

As regards the Primakoff effect (see Appendix):
\begin{equation}
\frac{d\sigma_{Prim}}{d \Omega} \simeq
\frac{g^2_{a \gamma \gamma}}{16 \pi^2}
sin^2(\theta)
m_a^3\; |\vec{k}| 
\left|
\int d^3x \Phi(\vec{x}) e^{i\vec{q} \cdot \vec{x}}
\right|^2
\label{eq:sfi5}
\end{equation}
where $\theta$ is the scattering angle between $a$ and $\gamma$, $\vec{k}$ is the $a$
momentum and $\Phi$ is the electric potential of the target nuclei.

Since, the NaI crystal is a ionic crystal, 
the electric potential generated by each ion 
is well approximated by two components, 
a Yukawa one shielded by the bound electrons and 
a Coulomb one at long range, respectively:
\begin{equation}
\Phi_{b}(\vec{x})=
\frac{(Z_b \pm 1)e}{4 \pi x} e^{-\frac{x} {r_b}}
\mp \frac{e}{4 \pi x}
\label{eq:pot1}
\end{equation}
where the upper(lower) sign is for $b=I$($b=Na$) and 
$r_b$ is the screening radius of the atom.
The Fourier transform of the electric potential is:
\begin{equation}
F_b(q)=
\frac{(Z_b \pm 1)e}{q^2+\frac{1} {r^2_b}} 
\mp \frac{e}{q^2}
\label{eq:pot2}
\end{equation}

Hence, the counting rate $R_{Prim}$ can be expressed as:

\begin{equation}
R_{Prim}=
\frac{N_T \rho_a m_a^3 g^2_{a \gamma \gamma}}{6 \pi}
\sum_{b=Na,I}
\left|
F_b(q)
\right|^2
\langle
v^2 
\rangle
\label{eq:rate3}
\end{equation}
with $q \simeq m_a$.

Therefore, also in this case, as
for the axioelectric one, 
a non-zero modulation term is present.

\subsubsection{Conclusions}

In conclusion, all the kinds of interaction discussed above for the pseudoscalar
candidate contribute to the costant part of the signal, while 
the Compton-like interaction does not contribute to the modulation part of the signal.
Moreover, as it can be easily demonstrated, for the pseudoscalar case 
the axioelectric contribution to the total expected counting rate is largely dominant
with the respect to the Primakoff and Compton-like on nuclei contributions at least in all the
``natural'' cases, where $g_{a\bar{e}e}/m_e$ is not lower than a factor $\sim 10^{-3}$ 
the coupling constant to mass ratios of the other charged fermions; 
in addition, it still remains at least one order of magnitude 
larger than the one due to the Compton-like effect
on electrons, for $a$ particle mass below $\simeq$ 6 keV.

\subsection{The data analysis and results for the DM pseudoscalar candidate} \label{ana1}

A complete data analysis of the 107731 kg $\cdot$ day exposure from DAMA/NaI
in this framework would offer an allowed multi-dimensional volume in the space
defined by $m_{a}$ and by all the $g_{a \bar{f}f}$ coupling constants to charged 
fermions.
Since a graphic representation 
would be practically impossible, we will present some particular slices 
of such an allowed volume.

In the following calculations, we have considered some of the possible uncertainties
in the parameters needed for the counting rate evaluations. In particular, the screening
radia of Sodium and Iodine atoms have been taken in the range: 
$\frac{1}{r_{Na}} = (2.0 \pm 0.5)$ keV and
$\frac{1}{r_{I}}  = (5.0 \pm 0.5)$ keV, respectively.
Moreover several halo models have been considered \cite{RNC,IJMPD} either
spherically symmetric matter density with isotropic velocity dispersion
or 
spherically symmetric matter density with non-isotropic velocity dispersion
or axisymmetric models or triaxial models.
The parameters of each halo model have been chosen taking into account the available
observational data; in particular the local velocity has been varied within its
allowed range: $v_0=(220 \pm 50)$ km s$^{-1}$ (90\% C.L.) and local density
$\rho_{halo}$ has been varied within the range evaluated as in ref. \cite{Hep}
taking into account the following physical constraints: i) the amount of flatness of the rotational
curve of our Galaxy, considering conservatively $0.8 \cdot v_0  \lsim
v_{rot}^{100} \lsim  1.2 \cdot v_0$, where $v_{rot}^{100}$ is the
value of rotational curve at distance of 100 kpc from the galactic
center; ii) the maximal non dark halo components in the Galaxy, considering
conservatively $1 \cdot 10^{10} M_{\odot}  \lsim  M_{vis}  \lsim  6
\cdot 10^{10} M_{\odot}$ \cite{Deh98,Gat96}.
Although  a large number
of self-consistent galactic halo models, in which the variation of the
velocity distribution function is originated from the change of the
halo density profile or of the potential, have been considered, still many other possibilities exist.
The parameter $\langle v_g^2 \rangle$ has been evaluated for each considered halo model
and for the given local velocity value.

The results presented in the following by using the DAMA/NaI annual modulation data
over the seven annual cycles are calculated by taking into
account the time and energy behaviours of the {\it single-hit} experimental data
through the standard maximum likelihood 
method\footnote{Shortly, the likelihood function is: 
${\it\bf L}  = {\bf \Pi}_{ijk} e^{-\mu_{ijk}}
{\mu_{ijk}^{N_{ijk}} \over N_{ijk}!}$, where
$N_{ijk}$ is the number of events collected in the
$i$-th time interval, by the $j$-th detector and in the
$k$-th energy bin. $N_{ijk}$ follows a Poissonian
distribution with expectation value 
$\mu_{ijk} = [b_{jk} + S_{0,k} + S_{m,k} \cdot  cos\omega(t_i-t_0)] M_j \Delta
t_i \Delta E \epsilon_{jk}$. 
The unmodulated and modulated parts of the signal,
$S_{0,k}$ and $S_{m,k}cos\omega(t_i-t_0)$, respectively,
are functions of the coupling constants and of the light boson mass;
b$_{jk}$ is the background contribution;
$\Delta t_i$ is the detector running time during the $i$-th time interval; 
$\epsilon_{jk}$ is the overall efficiency and $M_j$ is the detector mass.}.

In particular, the likelihood function
requires the agreement: i) of the expectations for the modulation part of the signal
with the measured modulated behaviour for each detector and for each energy bin; ii)
of the expectations for the unmodulated component of the signal with the respect
to the measured differential energy distribution.
In the following for simplicity, the results of these corollary quests for the
pseudoscalar candidate particles are presented in terms of allowed regions
obtained as superposition of the configurations corresponding
to likelihood function values {\it distant} more than $3\sigma$ from
the null hypothesis (absence of modulation) in each of the several
(but still a limited number) of frameworks.
Obviously, these results are not exhaustive of the many scenarios
(still possible at present level of knowledge)
for these and for other classes of candidates, such as the WIMPs we already 
deeply investigated \cite{Mod1,Mod2,Ext,Mod3,Sist,Sisd,Inel,Hep,RNC,IJMPD}.

\begin{figure} [!ht]
\centering
\vspace{-1.0cm}
\includegraphics[width=250pt] {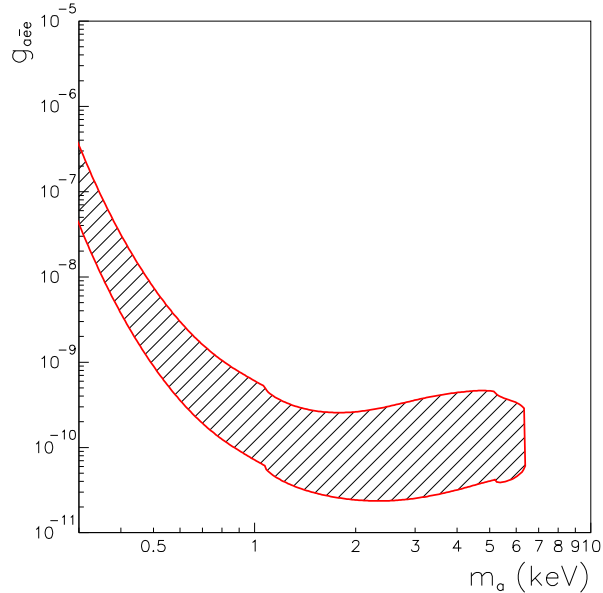}
\vspace{-0.8cm}
\caption{Pseudoscalar case: region in the plane $g_{a\bar{e}e}$ vs $m_a$ allowed by the DAMA/NaI annual modulation data 
in the considered model framework. See text.}
\label{fg:axgaee}
\end{figure}

First of all, as already mentioned, the axioelectric contribution is dominant with the respect 
to the Compton-like and Primakoff effects in all the ``natural'' cases; thus, 
the results can be presented in terms of only two variables $g_{a \bar{e}e}$ 
and $m_a$. The allowed region in the plane defined by these two variables 
has been calculated considering the DAMA/NaI results on the model independent annual modulation
signature and has been reported in Fig. \ref{fg:axgaee}.
It is worth to note that, for the reasons reported above, this allowed region is almost 
independent on the adopted $g_{a \bar{u}u}$ and $g_{a \bar{d}d}$ coupling constants.
The allowed region reported in Fig. \ref{fg:axgaee} can only marginally be
affected by the results already presented at low energy by low-background
ionization detectors;
in fact, due to their energy resolution and to their quoted counting rate at low energy, 
their results do not rule out $a$ particles with $m_a \lsim 3$ keV and,
for $m_a \gsim 3$ keV, $a$ particles with 
$g_{a\bar{e}e} \lsim 2 \times 10^{-10}$.

Some strongly model dependent astrophysical limits on the $g_{a\bar{e}e}$ 
can be found in literature (see e.g. \cite{raf95}) by
studying the globular cluster stars; however, these constraints only apply
to particles with masses much below few keV, which is the typical core temperature of the stars.

\begin{figure} [!ht]
\centering
\vspace{-0.4cm}
\includegraphics[width=310pt] {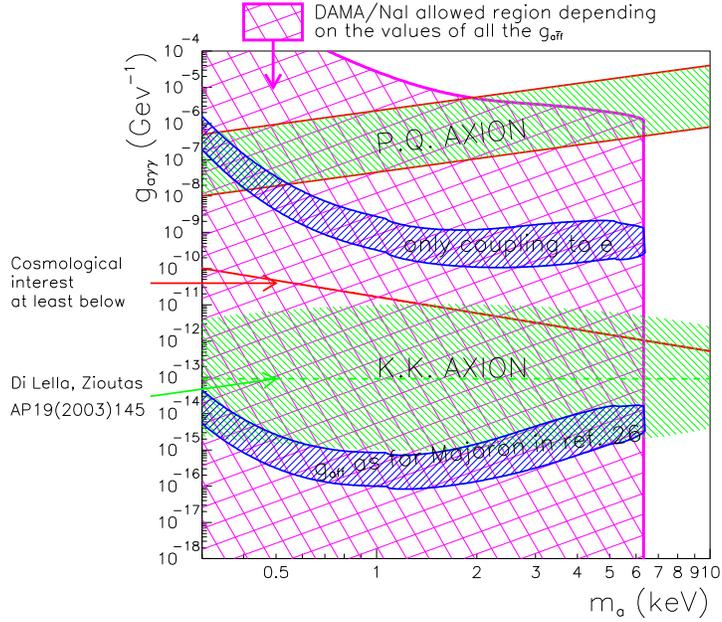}
\vspace{-0.4cm}
\caption{DAMA/NaI allowed region at 3$\sigma$ C.L. in the plane $g_{a\gamma\gamma}$ vs $m_a$
for a light pseudoscalar candidate ({\em crossed hatched region}). 
All the configurations in this region can be allowed 
depending on the values of all the $g_{a \bar{f}f}$.
Examples of two of the many possible models: i) {\em upper black region}: coupling only to electrons;
ii) {\em lower black region}: coupling (proportional to the $m_a$)
through the (weak) isospin to 
quarks and leptons. For details see text. 
The indicative region of the Kaluza-Klein (K.K.) pseudoscalar axion credited in ref. \cite{DiLella} from the analysis
of indirect observations and the region of the DFSZ and KSVZ models (P.Q. axion) are shown as well. 
The solid line corresponds to $a$ particle with lifetime equal to the
age of the Universe; at least all the $g_{a \gamma\gamma}$'s below this line are of cosmological interest.  
See text.
Thus, a pseudoscalar DM candidate could also account for the DAMA/NaI model independent result as
well as the WIMP solutions already discussed elsewhere.}
\label{fg:axeqq}
\end{figure}

Let us now investigate the cosmological interest of the allowed
$a$ particle, actually the case of an $a$ lifetime of order or larger than the age of the Universe.
As reported in eq. (\ref{eq:ta}), this lifetime is directly connected to the
$g_{a \gamma \gamma}$ value. 
Hence, we show -- for a general case -- in Fig. \ref{fg:axeqq} the DAMA/NaI region allowed in the "usual" plane 
$g_{a \gamma\gamma}$ vs $m_a$ (crossed hatched region).
The upper bound on $g_{a \gamma \gamma}$ is given when the Primakoff effect is largely dominant
(that is, if $g_{a \bar{e}e}=g_{a \bar{u}u}=g_{a \bar{d}d}=0$ and only contributions from other charged fermions
are present).
All the other values of $g_{a \gamma\gamma}$ below this upper bound are allowed
depending on the values of all the $g_{a \bar{f}f}$.
Just for example, two particular (of the many possible) sets of $g_{a \bar{f}f}$ are reported in Fig. \ref{fg:axeqq}
as {\em upper black region} and {\em lower black region}, respectively:
the case of the coupling only to electrons ($g_{a \gamma \gamma} \simeq
\frac{g_{a \bar{e}e}\alpha}{\pi m_e}$)
and the case of $g_{a \bar{u}u}$ and $g_{a \bar{d}d}$
according to the scenario of ref. \cite{RG} for the Majoron. 
This last model assumes that $g_{a \bar{f}f} \propto m_f$:
$\frac{ g_{a \bar{e}e} }{m_e} = \frac{ g_{a \bar{d}d} }{m_d} = - \frac{g_{a \bar{u}u}}{m_u}$;
same relations also hold for the other families.
Replacing these assumptions in the eq. (\ref{eq:triangolos}), we have at the first order:
\begin{equation}
g_{a \gamma \gamma} \sim \left[\frac{g_{a \bar{e}e}}{m_e}
+\frac{g_{a \bar{d}d}}{3m_d}
+\frac{4 g_{a \bar{u}u}}{3 m_u}\right] \sim \left[1+\frac{1}{3}-\frac{4}{3}\right]
\sim 0
\label{eq:triangolos3}
\end{equation}
while at the second order:
\begin{equation}
g_{a \gamma \gamma} \sim
\frac{g_{a \bar{e}e}\alpha}{12\pi m_e}
\frac{m^2_a} {m^2_e} \simeq 1.4 \; 10^{-6} \; GeV^{-1} \cdot g_{a \bar{e}e} \cdot \left( \frac{m_a} {1 \; keV} \right)^2 .
\label{eq:triangolos4}
\end{equation}

Fig. \ref{fg:axeqq} also shows 
the result of Kaluza-Klein axions credited in ref. \cite{DiLella} from analysis 
of indirect observations; it is well embedded 
in the DAMA/NaI allowed region. Moreover,  
when considering the existing uncertainties (qualitatively reported in figure as shaded area), 
the result of ref. \cite{DiLella} can also be in agreement with the DAMA/NaI region allowed for the
scenario of ref. \cite{RG} depicted in the {\em lower black region}.
Moreover, the DAMA/NaI allowed region is also comparable with the predictions of ref. \cite{Gorb01} 
of axion-like particles as UHECR.

The solid line in Fig. \ref{fg:axeqq} corresponds to $a$ particle with lifetime equal to the 
age of the Universe; therefore, at least all the configurations with $g_{a \gamma\gamma}$ below this line are of 
cosmological interest. Moreover, in principle, it might be possible that the configurations above 
this line would also become of interest in case of some exotic mechanism of the $a$ particle production.

Thus, a pseudoscalar DM candidate could also account for the DAMA/NaI model independent result as
well as the WIMP solution for several (but still few with the respect to the possibilities) 
corollary model dependent quests for the candidate particle
already discussed elsewhere \cite{Mod1,Mod2,Ext,Mod3,Sist,Sisd,Inel,Hep,RNC,IJMPD}.

\section{Scalar case}

The effective Lagrangian of interaction of the scalar particle
(we remind that here and hereafter the scalar particle is named $h$)
is given by the sum of the Yukawa couplings to the fermions
and the effective coupling to photons through a charged fermion loop:

\begin{equation}
L_{int} = g_{h \bar{f}f} h\; \bar{f} f + 
\frac{g_{h \gamma \gamma}}{4} h \; F_{\mu \nu}F^{\mu
\nu}
\label{eq:lhff}
\end{equation}

The corresponding diagrams are similar to those reported
in Fig. \ref{fg:diagramf}, just replacing $a$ with $h$. 
Also in this case the scalar $h$ particle can decay into photons
with lifetime $\tau_h = \frac {64 \pi} {g_{h \gamma \gamma}^2 m_h^3}$;
as regards the decay into pair of fermions 
similar considerations of sec. \ref{sec3} still hold.

The effective coupling constant to two photons $g_{h \gamma \gamma}$ is for
$m_h << m_f$ \cite{coupl00}:
\begin{equation}
g_{h \gamma \gamma} = \sum_f \frac{g_{h \bar{f}f} Q^2_f \alpha}{\pi\;m_f}
\left[-\frac{2}{3}-
\frac{7}{180}
\frac{m_h^2}{m_f^2}+O \left(\frac{m_h^4}{m_f^4} \right)
\right]
\label{eq:hloop3}
\end{equation}

\subsection{Direct detection of the scalar particle}

\subsubsection{The Compton-like effect}

The differential cross section of the Compton-like effect 
(see also Fig.\ref{fg:diagr}) for $h$ particle impinging on 
point-like free fermion is given by the sum of two contributions
(see the Appendix):
\begin{equation}
\frac{d\sigma_{C-l,f}}{d\Omega} \simeq 
\frac{\alpha Q_f^2 g^2_{h \bar{f}f} m_h^2}   {8\pi\;m_f^4 |\vec{v}|}
\left[1+2\frac{m_f^2}{m_h^2}|\vec{v}|^2 sin^2(\theta)\right]
\label{eq:ssig1}
\end{equation}
The first term is similar than that obtained for the pseudoscalar
interaction and, since it is proportional to the inverse of $h$ velocity,
it does not contribute to the modulation amplitude of the signal;
the second term is instead proportional to 
$h$ velocity and it does contribute to the modulation part of the signal.
   
The expected counting rate, $R_{C-l}$, in a detector can be written
as the sum of $R_{C-l,elect}$ and $R_{C-l,nucl}$ given by the 
conversion on the atomic electrons 
and by the conversion on the nuclei, respectively.

For the conversion on electrons:
\begin{equation}
R_{C-l,e} = N_T \sum_{A=Na,I} \frac{\rho_a \alpha Z_{A}^2 g^2_{h \bar{e}e} m_h} {2 m_e^4} 
\left[1+\frac{4}{3}\frac{m_e^2}{m_h^2} \langle v^2 \rangle \right]  F_{A,el}^2(q)
\label{eq:ratecomh1}
\end{equation}

For the nucleus:
\begin{equation}
R_{C-l,nucl} = N_T \sum_{A=Na,I} \frac{\rho_a \alpha Z_{A,eff}^2 g^2_{h \bar{A}A} m_h} {2 m_A^4} 
\left[1+\frac{4}{3}\frac{m_A^2}{m_h^2} \langle v^2 \rangle \right]  F^2_{A,Nucl}(q)
\label{eq:ratecomh2}
\end{equation}

The scaling law for the 
$g_{h \bar{A}A}$ has been evaluated considering the coherent 
contribution of all the quarks: $
g_{h \bar{A}A} = A \left[ (g_{h \bar{u}u} + 2 g_{h \bar{d}d})
+ \frac{Z}{A} (g_{h \bar{u}u} - g_{h \bar{d}d}) \right] $.
Since the ratio $Z/A$ is quite similar for the nuclei (0.48 and 0.42 for the $^{23}$Na 
and for the $^{127}$I, respectively), 
we can write without losing generality 
$ g_{h \bar{A}A} = A g_{h \bar{N}N} $, where 
$ g_{h \bar{N}N} $ is the effective coupling constant to a single nucleon.
The effective charge of nuclei, $Z_{A,eff}$, takes into account the 
screening of the inner electrons; a reasonable approximation for
$m_h \sim$ keV is $Z_{Na,eff} = 9$ and $Z_{I,eff} = 35$
(see Tab. \ref{tb:shell}).
The form factors for the nuclei can be considered as one since the nuclei 
are practically point-like for the energy of interest. 

Finally, the Compton-like effect provides in the present case 
a counting rate varying along the year.

\subsubsection{The photoelectric-like effect}

The cross section of the photoelectric-like effect of a scalar $h$ particle
can be evaluated as reported in the Appendix and is:
\begin{equation}
\sigma_{p-e,nl} = 
g^2_{h \bar{e}e}
\frac{2 \pi  m_e} {|\vec{k}|} |\vec{p}| 
F^2_{nl}(q)
\label{eq:FGR2}
\end{equation}
The counting rate $R_{p-e}$ can be calculated straighforward as done 
in sec. \ref{a-e} for the pseudoscalar case.
\begin{equation}
R_{p-e} =
N_T
\frac{\rho_a 2\pi g^2_{h\bar{e}e} m_e} {m_h^2}
\sum_{b=Na,I} \sum_{nl} N_{nl}
F^2_{nl}(q) \; p \;
\Theta(m_h-E_{nl})
\label{eq:sratecomae23}
\end{equation}
It is worth to note that the photoelectric-like effect for the scalar case
gives contribution only to $S_0$ part of the signal and does not
provide any significant time dependent contribution.

\subsubsection{The Primakoff effect}

The cross section and the expected counting rate for
the Primakoff effect are equal than those reported 
for the pseudoscalar case, with the obvious substitution
of the coupling constant (see Appendix).
As in that case, a non-zero modulation term is expected.

\subsubsection{Conclusions}

In conclusion, all the kinds of interactions discussed above for the scalar
candidate contribute to the costant part of the signal, while 
the photoelectric-like interaction does not contribute to the modulation part of the signal.
However, the photoelectric-like interaction gives a dominant contribution with the
respect to the Compton-like process on the electrons and to the other effects.
Therefore, the coupling to the electrons
does not produce any significant time variation of the signal. Thus,
in the following we will investigate the case of a scalar $h$ particle 
coupled only to the hadronic matter;
in this case, non-zero modulation term of the signal is expected by 
the contributions of the Compton-like effect on nuclei and of the Primakoff effect.

\subsection{The data analysis and results for the DM scalar candidate}

\begin{figure} [!ht]
\centering
\vspace{-0.9cm}
\includegraphics[width=200pt] {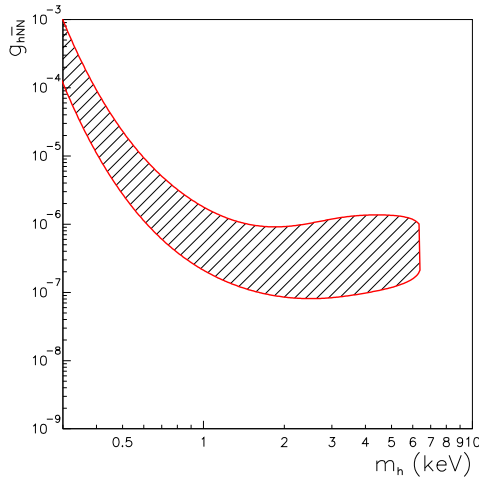}
\vspace{-0.6cm}
\caption{DAMA/NaI allowed region at 3$\sigma$ C.L. in the plane $g_{h\bar{N}N}$ vs $m_h$
for a scalar light boson DM candidate coupled only to the hadronic matter
in the given frameworks. To account for cosmological interesting lifetimes,
here only the configurations with negligible Primakoff effect contribution 
have been selected. See text.}
\label{fg:quark}
\end{figure}

Processes involving DM scalar candidate coupled to electrons do not contribute to the
annual modulation of the signal. Moreover, the upper limit on $g_{h\bar{e}e}$, which can be easily derived 
from the energy distribution measured by DAMA/NaI (given elsewhere), 
ranges from $\simeq 3 \times 10^{-16}$ to $\simeq 10^{-14}$ for $m_h$
between $\simeq 0.5$ and 10 keV. 
Thus, in the following we investigate the 
case of a scalar $h$ particle coupled only to the hadronic matter
($g_{h\bar{e}e} << g_{h\bar{u}u}, g_{h\bar{d}d}$).
This framework would offer an allowed multi-dimensional volume in the space
defined by $m_{h}$ and by all the $g_{h \bar{f}f}$ coupling constants to quarks.
However, to account for cosmological interesting $h$ particle lifetimes,
only the configurations with negligible Primakoff effect contribution can be
investigated and only the diagrams at the tree level with quarks are considered.
Therefore, in this case
the Compton-like effect on nuclei is 
the major contribution to the cross section. 
Hence, according to the scenarios already described in sec. \ref{ana1},
we report in Fig. \ref{fg:quark} the region allowed by the DAMA/NaI data 
in the plane $g_{h \bar{N}N}$ vs $m_h$. 
This allowed region can only marginally be 
affected by the results already presented at low energy by low-background
ionization detectors,
which do not rule out $h$ particles with $m_h \lsim 3$ keV and,
for $m_h \gsim 3$ keV, $h$ particles with 
$g_{h\bar{N}N} \lsim 8 \times 10^{-7}$.
Finally, the strongly model dependent 
globular cluster constraints \cite{pdg} 
only apply to particles with masses much below few keV.

\begin{figure} [!ht]
\centering
\vspace{-0.4cm}
\includegraphics[width=200pt] {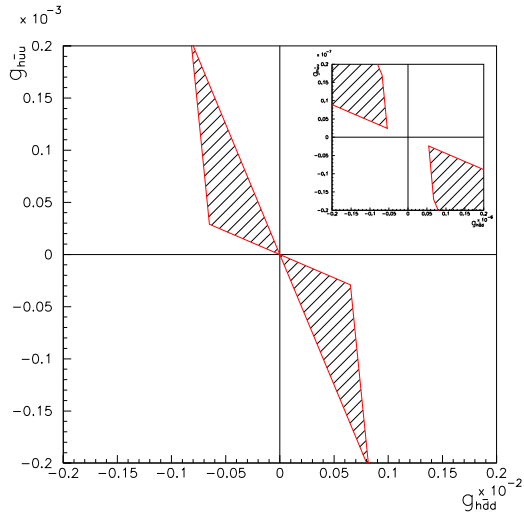}
\vspace{-0.6cm}
\caption{DAMA/NaI allowed configurations of cosmological interest
in the $g_{h \bar{u}u}$ vs $g_{h \bar{d}d}$ plane
obtained by
requiring that they: i) are allowed in the plane $g_{h\bar{N}N}$ vs $m_h$
in Fig. \ref{fg:quark}; ii) $m_h > 0.3$ keV; 
iii) $\tau_h \gsim$ the age of the Universe. The uncertainties on the first family quark
masses have also been accounted for:
$m_u \sim 3.0 \pm 1.5$ MeV, $m_d \sim 6.5 \pm 2.0$ MeV. 
In the inset: a magnification of the regions around (0,0).
See text.
Thus, also a scalar DM candidate could account for the DAMA/NaI model
independent result as the previous case of pseudoscalar DM candidate and 
the WIMP solutions already discussed elsewhere.}
\vspace{-0.4cm}
\label{fg:liveq}
\end{figure}

Obviously, the region in Fig. \ref{fg:quark} does not allow a direct information
about the effective coupling constant to photons, $g_{h\gamma\gamma}$,
and, therefore, about $h$ lifetime; however, the large number of free coupling constants 
allows to expect the existence of a large number of $h$ configurations of cosmological interest.
For example, assuming as reasonable hypothesis that all the $g_{h \bar{q}q}$'s would be of the same
order of magnitude, $\tau_h$ would be dominated by the quarks of the first family
(see eq. (\ref{eq:hloop3})).
In this particular case, we can report in the $g_{h \bar{u}u}$ vs $g_{h \bar{d}d}$ plane 
the configurations of cosmological interest allowed by the results of DAMA/NaI.
Fig. \ref{fg:liveq} shows the shaded regions obtained by 
requiring that the configurations: i) are allowed in the plane $g_{h\bar{N}N}$ vs $m_h$
as reported in Fig. \ref{fg:quark}; ii) $m_h > 0.3$ keV; iii) are of a cosmological interest 
($\tau_h \gsim$ the age of the Universe). The uncertainties on the first family quark 
masses have also been accounted for \cite{pdg}:
$m_u \sim 3.0 \pm 1.5$ MeV, $m_d \sim 6.5 \pm 2.0$ MeV.

It is worth to note that we have not included in the presented calculation 
a possible contribution due to the scalar interaction of $h$ particle to the gluons, 
which can naturally explain the pure coupling to the nuclear matter. 
In principle, this contribution might increase the configurations of cosmological interest and
allowed by DAMA/NaI.

Finally, also a scalar DM candidate could account for the DAMA/NaI model
independent result as the previous case of pseudoscalar DM candidate and 
the WIMP solutions already discussed elsewhere.

\section{Conclusions}

In this paper we have discussed in some details another class of DM candidate particles: 
the pseudoscalar and scalar light bosonic candidates; particular care has been devoted
to the study of the processes for their detection in a suitable underground experimental set-up.
In future we also plan to explore additional scenarios which can -- in principle -- exist 
such as a particle with mixed pseudoscalar and scalar 
interaction, a triplet of (pseudo-)scalar
particles (which can ``naturally'' allow a quasi-stable bosonic candidate), 
particles with gluon coupling, spin 1/2 light dark matter candidates 
such as those of ref. \cite{Fayet,Ahluw}, etc..

Moreover, it is worth to explore also the possibility that (electron coupled) pseudoscalar $a$ particles 
produced by high energy sources in the galactic bulge would -- by interaction either with interstellar 
medium or with relic $a$ particle --
also contribute to positron production and to the 511 keV annihilation $\gamma$ rays from the bulge of Galaxy
observed in ref. \cite{integ}.
Finally, just candidates produced in the early Universe and decoupled from the primordial
Universe have been considered here; other mechanisms of light bosons production can 
in principle be considered, such as their production
by a heavier particles decays or ``exotic'' cosmological evolution of the Universe, etc..

In conclusion this paper, far to be exhaustive on the topics
(e.g. many other scenarios can be considered for the galactic halo), has however
shown that -- in addition to the WIMP cases already discussed by DAMA collaboration elsewhere 
(\cite{RNC,IJMPD} and references therein) -- there is also possibility for a bosonic candidate
with axion-like phenomenology and with mass $\lsim$ 6-7 keV to account for the 
6.3 $\sigma$ C.L. model independent evidence for the presence of a particle DM component in 
the galactic halo observed by DAMA/NaI.

The new higher sensitive DAMA/LIBRA set-up now in operation deep underground at
the Gran Sasso National Laboratory
of I.N.F.N. will allow to
restrict the possibility in the corollary
quests for the DM candidate particle as soon as it will have collected a well competing exposure with the
respect to the previous DAMA/NaI.

\section{Appendix}

\subsection{Compton-like cross section}

\subsubsection{Pseudoscalar case}

Let us consider the diagram of Fig. \ref{fg:diagr} where
a pseudoscalar $a$ particle with quadrimpulse $k$ is absorbed by
a charged fermion of quadrimpulse $p$ with the subsequent emission of a photon
with quadrimpulse $k'$. The outgoing fermion has quadrimpulse $p'$.
The matrix element for Compton-like conversion is given by the sum of
the contributions of this diagram
and of the "crossed" one. We get:
\begin{equation}
M  = i e Q_f g_{a \bar{f}f} 
\bar{U}(p') 
\left[ \not \!\epsilon
\frac{\not \!p + \not \!k + m_f}{(k+p)^2-m_f^2}
\gamma_5  
+ \gamma_5
\frac{\not \!p - \not \!k' + m_f}{(p-k')^2-m_f^2}
\not \!\epsilon 
\right] U(p) 
\label{eq:rCOM1}
\end{equation}
Using the standard algebra of the $\gamma$ matrices, the square of matrix 
element averaged on the initial fermion polarizations
and summed on the final electron polarizations can be written as:
\begin{equation}
\resizebox{0.85\textwidth}{!}{$
\overline{|M|^2} = 
\frac{-e^2 Q_f^2 g^2_{a \bar{f}f}} {2} 
Tr \left[(\not \!p'+m_f) \left(
\frac{\gamma_5 \not \,\epsilon \not \,k}
{m_a^2+2p \cdot k} 
- \frac{\gamma_5 \!
\not \,\epsilon \not \,k'}{2p \cdot k'} \right)
(\not \!p+m_f) 
\left(
\frac{\not \,k \not \,\epsilon
\gamma_5}
{m_a^2+2p \cdot k} 
- \frac{\not \,k' \! \not \,\epsilon
\gamma_5}{2p \cdot k'} \right) \right]
$}
\label{eq:rCOM4}
\end{equation}
After some algebric manipulations and
summing on the outgoing photon polarizations, we obtain:
\begin{equation}
\overline{|M|^2} =e^2 Q_f^2 g^2_{a \bar{f}f}
\left[2\;A+B|\vec{k}|^2 sin^2(\theta)\right]
\label{eq:rCOM10}
\end{equation}
where:
\begin{equation}
A = \frac{2k' \cdot p }
{m_a^2+2p \cdot k}  
 + \frac{2 \; p \cdot k+m_a^2}
{2 \;p \cdot k'} -2	
\label{eq:rCOM7}
\end{equation}
and 
\begin{equation}
B =- \frac{4}
{m_a^2+2p \cdot k}
+ \frac{
8 (k \cdot p) }
{(m_a^2+2p \cdot k)^2} 
\label{eq:rCOM8}
\end{equation}
The cross section is given by (see for example \cite{leaderpredazzi})
\begin{equation}
d\sigma = \frac{ \overline{|M|^2} }{|v_1-v_2|}
\frac{1}{2p_0}\frac{1}{2k_0}
\frac{d^3k'}{(2\pi)^3 2\;k_0'} 
\frac{d^3p'}{(2\pi)^3 2 \; p_0'} 
(2\pi)^4 \delta^4(p'+k'-p-k)
\label{eq:rsig1}
\end{equation}
For $m_f>>m_a$ and considering that both the $a$ particle and the fermion 
can be treated in 
the non-relativistic limit ($E_{\gamma} \simeq E_a \simeq m_a$), we obtain
$A \rightarrow \frac{m_a^2}{m_f^2}$ and $B |\vec{k}|^2 \rightarrow -\frac{|\vec{k}|^2}{m_f^2}
\simeq -\frac{m_a^2}{m_f^2} \beta^2$ ($ \beta \simeq 10^{-3}$ is the $a$ particle velocity
in the galactic halo).
Therefore, the cross section can be evaluated by considering that the $A$ term is dominant
in the squared matrix element:
\begin{equation}
\frac{d\sigma}{d\Omega \;dE_{\gamma}} =
\frac{\alpha Q_f^2 g^2_{a \bar{f}f}}     {m_f \pi}
\frac{m_a^3}                           {m_f^3}
\frac{1}                               {8 |\vec{k}|}
\delta(E_{\gamma}-E_a)
\label{eq:rsig3}
\end{equation}
and eq. (\ref{eq:sig3}) can be inferred.

\subsubsection{Scalar case}

The matrix element for Compton-like conversion for scalar case can be written,
using the same formulation as in the previous case, as:
\begin{equation}
M = e Q_f g_{h \bar{f}f} 
\bar{U}(p') 
\left[ \not \!\epsilon
\frac{\not \!p + \not \!k + m_f}{(k+p)^2-m_f^2} 
+
\frac{\not \!p - \not \!k' + m_f}{(p-k')^2-m_f^2}
\not \!\epsilon 
\right] U(p) 
\label{eq:rscom1}
\end{equation}
and the square of matrix 
element averaged on the initial fermion polarizations
and summed on the final electron polarizations:
\begin{equation}
\resizebox{0.85\textwidth}{!}{$
\overline{|M|^2} = 
\frac{e^2 Q_f^2 g^2_{h \bar{f}f}} {2} 
Tr \left[(\not \!p'+m_f) \not \!\epsilon  \left(
\frac{\not \,k + 2m_f}
{m_h^2+2p \cdot k} 
- \frac{\not \,k'}{2p \cdot k'} \right)
(\not \!p+m_f) 
\left(
\frac{\not \,k +2m_f}
{m_h^2+2p \cdot k} 
- \frac{\not \,k' }{2p \cdot k'} \right) \not \!\epsilon \right]
$}
\label{eq:rCOM42}
\end{equation}
After some algebric reductions and
summing on the outgoing photon polarizations, we obtain:
\begin{equation}
\overline{|M|^2} = e^2 Q_f^2 g^2_{h \bar{f}f}
\left[2\;A+C|\vec{k}|^2 sin^2(\theta)\right]
\label{eq:rSCOM6}
\end{equation}
where: $C = B + \frac{16 m_f^2}{(m_h^2+2p \cdot k)^2}$.
For $m_f>>m_a$ and in the non-relativistic limit:
$C |\vec{k}|^2 \rightarrow \frac{4 |\vec{k}|^2}{m_a^2}
= 4 \beta^2$.
Hence, on the contrary of the pseudoscalar case, both the $A$ and $C$ terms in the squared matrix
element can contribute at different extent to the total cross sections depending on
the $h$ particle mass and on the considered fermion, either electrons or nucleons.
Therefore, the non-relativistic cross section is:
\begin{equation}
\frac{d\sigma}{dE_{\gamma}d\Omega} = 
\frac{\alpha Q_f^2 g^2_{h \bar{f}f}}   {8\pi \beta m_f^2}
\left[
\frac{m_h^2 }{m_f^2}+2\beta^2 sin^2(\theta)
\right]
\delta(E_{\gamma}-E_h)
\label{eq:rssig1}
\end{equation}
from which eq. (\ref{eq:ssig1}) can be derived.

\subsection{Photoelectric-like cross section}

\subsubsection{Pseudoscalar case, axioelectric process}

From the Lagrangian (\ref{eq:laff}) it is possible to obtain in the non-relativistic limit 
the Schroedinger-Pauli equation for the electron in presence of the 
$a$ field and of the electromagnetic field:
\begin{equation}
\left( i\partial_{0} + e A_{0} \right) \psi
 + \frac{\left(\vec{\nabla} + ie\vec{A}\right)^2}{2m_e}
\psi
- \frac{e}{2m_e}\vec{\sigma} \cdot
(\vec{\nabla} \times \vec{A})
\psi
-g_{a \bar{e}e}
\frac{\vec{\sigma} \cdot   
\vec{\nabla}(a)}{2m_e} \psi = 0
\label{eq:alele6}
\end{equation}
It is evident, therefore, as in the non-relativistic limit, the $a$ particle interaction 
to fermions is spin-dependent.

The cross section of the process can be evaluated through the Fermi golden rule:
\begin{equation}
d\sigma = \frac{W_{if}}{J_i}= 2\pi \frac{|M_{if}|^2}{J_i} \rho_f \; ,
\label{eq:raFGR}
\end{equation}
when considering
the perturbation term in the electron Hamiltonian (see for example \cite{Dimopoulos}), 
$V_a = g_{a \bar{e}e}\frac{\vec{\sigma} \cdot
\vec{\nabla}(a)}{2m_e}$.

Considering the $a$ field and the wavefunction of the escaping electron 
as plane waves and the wavefunction of initial
bounded electron in the $nlm$ shell as $\psi_{n l m}$, the transition
matrix element can be written as:
\begin{equation}
|M_{if}| = \left| \int \frac{e^{-i\vec{p} \cdot \vec{x}}}{\sqrt{V}} 
g_{a \bar{e}e} a_0 e^{i \vec{k} \cdot \vec{x}}
\frac{\vec{\sigma} \cdot
\vec{k}}{2m_e}
\psi_{n l m} d^3x \right|
\end{equation}
The incident flux is: $J_i = \left| -i (a^* \vec{\nabla} a - a \vec{\nabla} a^*) \right| = 2 a_0^2 k$ 
and the density of states for the outgoing electron is:
\begin{equation}
\rho_f = \frac{Vp^2dp}{(2\pi)^3} d\Omega 
\delta(E_a-E_{n l}-E_e)=\frac{m_eVp\;dE_e}{(2\pi)^3} d\Omega \delta(E_a-E_{n l}-E_e)
\label{eq:radens}
\end{equation}
where $E_{n l}$ is the binding energy of the atomic electron in the $nl$ shell.
Summing over the final electron polarizations and averaging on the initial
polarizations, we get:
\begin{equation}
\frac{d\sigma}{dE_e d\Omega}=
g^2_{a\bar{e}e}
\frac{k}{4m_e}
\left| \int e^{-i( \vec{p} - \vec{k} ) \vec{x}} \psi_{nlm} d^3x \right|^2
\frac{p}{(2\pi)^2} \delta(E_a-E_{nl}-E_e)
\label{eq:raFGR2}
\end{equation}
From this formula, eq. (\ref{eq:aFGR2}) can be deduced straighforward.

\subsubsection{Scalar case}

The procedure to evaluate the cross section for the scalar photoelectric-like 
case is similar to that applied before for the pseudoscalar case.
Considering the Lagrangian given in eq. (\ref{eq:lhff}),
the last term of the Schroedinger-Pauli equation (\ref{eq:alele6})
has to be replaced by the term $g_{h \bar{e}e} h \psi = -V_h \psi$.
Therefore, the cross section for the process in the non relativistic
limit can be written as:
\begin{equation}
\frac{d\sigma}{dE_e d\Omega}=
g^2_{h\bar{e}e}
\frac{m_e} {k}
\left| \int e^{-i( \vec{p} - \vec{k} ) \vec{x}} \psi_{nlm} d^3x \right|^2
\frac{p}{(2\pi)^2} \delta(E_h-E_{nl}-E_e)
\label{eq:rrFGR2}
\end{equation}
and the eq. (\ref{eq:FGR2}) can be inferred.

\subsection{Primakoff cross section}

\subsubsection{Pseudoscalar case}

Let us now consider the case of the Primakoff conversion 
of the pseudoscalar $a$ particle of momentum 
$\vec{k}$ into a photon of momentum $\vec{p}$ in the static
atomic field $F_{\rho \sigma}(\vec{x})$.
The S-matrix element is:
\begin{equation}
\left| S_{fi} \right|^2 = \left| \int d^4x 
\frac{2 g_{a \gamma \gamma}}{8} a \; \varepsilon ^ {\mu \nu \rho 
\sigma} F_{\mu \nu} 
F_{\rho \sigma}(\vec{x}) \right|^2
\label{eq:sfi}
\end{equation}
where the factor 2 takes into account the two photon crossed diagrams. 
Let us now consider the $a$ field and the wavefunction of the outgoing photon 
as plane waves: 
$a = a_0 e^{-ik \cdot x}$
and
$A_{\mu} = \epsilon_{\mu} \frac{e^{ip \cdot x}}
{\sqrt{2\;p\;V}}$.
For the static electric field: $\varepsilon ^ {\mu \nu \rho \sigma}
F_{\rho \sigma}(\vec{x}) = 2 \varepsilon ^ {\mu \nu 0 i} E_i(\vec{x}) =
-2 \varepsilon ^ {\mu \nu 0 i}\nabla_i \Phi(\vec{x})$.

The cross section is given by (see for example \cite{Bjorken}):
\begin{equation}
d\sigma = V \frac{d^3p}{(2 \pi)^3} \frac{\vert S_{fi} \vert^2}{T \;J_i}
\end{equation}
As for the previous case, the incident flux is given by: 
$J_i = \left| -i (a^* \vec{\nabla} a - a \vec{\nabla} a^*) \right| = 2 a_0^2 k$ 
and the time integral for a transition
in the time T can be obtained following the prescriptions given
in ref. \cite{Bjorken}.
Summing on the possible polarization of the outgoing photon,
we obtain the differential cross section of the process:
\begin{equation}
\frac{d\sigma}{d \Omega dE_{\gamma}}=
\frac{g^2_{a \gamma \gamma}}{16 \pi^2}
sin^2(\theta)
E_{\gamma}^3\;k 
\left|
\int d^3x \Phi(\vec{x}) e^{i\vec{q} \cdot \vec{x}}
\right|^2
\delta(E_{\gamma}-E_a)
\label{eq:rsfi5}
\end{equation}
Similar results have been obtained in ref. \cite{sikivie2}. 
It is worth to note that this case also includes the ultrarelativistic limit,
considered for example for the solar axion investigations.
Eq. (\ref{eq:sfi5}) can be directly achieved by the previous eq. (\ref{eq:rsfi5}).

\subsubsection{Scalar case}

For a scalar $h$ particle the S-matrix element for
Primakoff conversion is:
\begin{equation}
\left| S_{fi} \right|^2 = \left| \int d^4x
\frac{2 g_{h \gamma \gamma}}{4} h \;
F_{\mu \nu}
F^{\mu \nu}(\vec{x}) \right|^2
\label{eq:rssfi}
\end{equation}
Using the same procedure as for the pseudoscalar case
and reminding that the electric static field is:
\begin{equation}
F^{\mu \nu}(\vec{x}) = 
- \left[ \eta^{\mu 0}\eta^{i \nu}-\eta^{\mu i}\eta^{0 \nu} \right]\nabla_i \Phi
\label{eq:rfors}
\end{equation}
one can obtain the differential cross section as for the 
pseudoscalar case: see eq. (\ref{eq:rsfi5}).

\end{document}